\documentclass{iau} 
\usepackage{graphicx}
\usepackage{amsmath}
\usepackage{textcomp}
\usepackage{amssymb}
\usepackage{float}
\usepackage[colorlinks=true,citecolor=blue]{hyperref}

\title[Kodaikanal Digitized WL Data Archive] 
{An Overview of Science Results Obtained From Kodaikanal Digitized White-Light Data Archive: 1921-2011}

\author[Sudip Mandal \etal]   
{Sudip Mandal $^1$
 \and Dipankar Banerjee$^{1,2}$}

\affiliation{$^1$Indian Institute of Astrophysics, \\ Banglore 560034, India
\\ email: {\tt sudip@iiap.res.in} \\[\affilskip]
$^2$Center of Excellence in Space Sciences India, IISER Kolkata,\\ Mohanpur 741246, West Bengal, India
 \\email: {\tt dipu@iiap.res.in}}

\pubyear{2018}
\volume{340}  
\jname{Long-term datasets for the understanding of solar and stellar magnetic cycles}
\editors{A.C. Editor, B.D. Editor \& C.E. Editor, eds.}
\begin{document}

\maketitle
\begin{abstract}
 In this proceeding, we present a summary of the recent scientific results that have been derived using the newly digitized whit-light (WL) data obtained from the Kodaikanal Solar Observatory.

\keywords{Sun: sunspots , Sun: magnetic fields}
\end{abstract}

\firstsection 
\vspace{-0.3cm}             
\section{Introduction}

Full disc images of the Sun, in white-light (WL), are being recorded at the Kodaikanal Solar Observatory since 1904. Photographic plates/films, on which these images were originally recorded , have now been digitized in a high resolution 4K$\times$4K format, for a period of 90 years (1921-2011). Calibration of this newly digitized data is also complete now and various steps involved in the calibration process are described in details in \cite{2013A&A...550A..19R}. 

\vspace*{-0.5 cm}
\section{Sunspot Area Time Series and Its Analysis}
\begin{figure*}[!htb]
\centering
\includegraphics[width=0.98\textwidth]{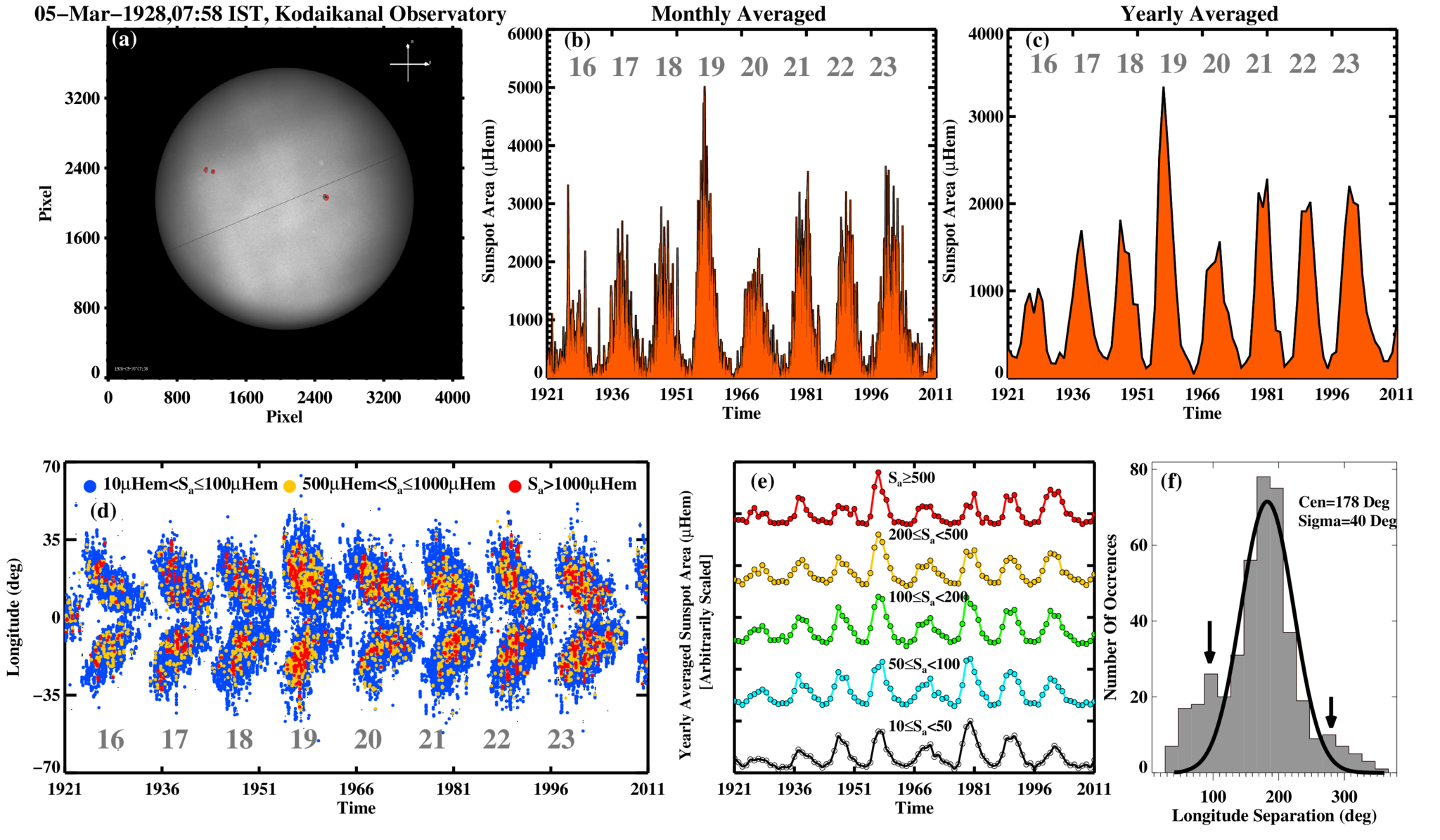}
\caption{A collage of the representative examples. Panel-(a) shows a Kodaikanal full disc WL digitized (and calibrated) image, with red coloured contours highlighting the detected Sunspots. Monthly and yearly averaged sunspot area cycles are presented in panels-(b-c). Panel-(d) shows the sunspot `butterfly diagram' whereas panel-(e) contains the individual area curves computed by considering certain sunspot size ranges. A histogram of separation, between two most active longitudes, is presented in panel-(f). } 
\label{fig1}
\end{figure*}

To identify the sunspots, we have used a semi-automated sunspot detection algorithm, optimized for Kodaikanal data in-order to handle multiple artefacts and varying image contrast (\cite{2013A&A...550A..19R}). For every detected spots, we record their heliographic positions and area (foreshortening corrected) . Using this algorithm, \cite{2017A&A...601A.106M} have built a sunspot area time series of 90 years (1921-2011) from the Kodaikanal WL data (Panels~\ref{fig1}.(b-c)). Various other properties of this time series, including the butterfly diagram, north-south asymmetry, Waldimier effect and multi-scale periodicities, are also studied in-detail in  \cite{2017A&A...601A.106M}.

As seen from the butterfly diagram (Fig~\ref{fig1}.d), the small spots are ubiquitous in time whereas the bigger spots mostly appear near the cycle maxima. Following this up, we perform a statistical study on the relationship between sunspot sizes and the solar cycle in \cite{2016ApJ...830L..33M}. We separately calculate the yearly averaged sunspot area for five defined spot size ranges (Fig~\ref{fig1}.e) and notice that though the overall cycle behaviours are primarily governed by the bigger spots, individual size ranges may not necessarily follow the same trend. An interesting example is the case of cycle 19 and cycle 21. Considering the overall amplitudes, cycle 19 is considerably stronger than cycle 21 whereas the smallest spot areas (10$\mu$Hem$\leq$50$\mu$Hem) in cycle 21 seems to be significantly higher than that of cycle 19. This result indicates that smaller spots may have a different generation mechanism as compared to the bigger spots. Other features of these individual area curves, such as even-odd rule, double peaks and north-south asymmetries, are also presented in \cite{2016ApJ...830L..33M}.

In a follow-up work, we investigate the `active longitude' problem using these detected sunspots and the results are presented in \cite{2017ApJ...835...62M}. From our analysis, we find a persistent presence of active longitudes during all the 8 solar cycles (cycles 16-23) studied in this case. Interestingly we also find that the separation between the two most active longitudes is roughly 180\textdegree~ with weaker presence of separations at 90\textdegree~and 270\textdegree (Fig~\ref{fig1}.f). The fact that these separations occur in multiples of 90, is very intriguing and needs further investigations. Apart from this, we also note that the active longitudes show a `flip-flop' like phenomena with a period of $\approx$2 years which has also been observed previously in many other stars (\cite{2007MmSAI..78..242B}).

\vspace*{-0.3 cm}
\section{Conclusion}
In this short article, we provide a summary about the newly digitized and calibrated white-light data obtained from Kodaikanal solar observatory for a period of 90 years. We have also highlighted some key results from three of our recent works based on this data. This data is now available for public download at \url{https://kso.iiap.res.in/new}.
\vspace{-0.4cm}


\end{document}